\documentclass[a4paper,11pt]{article}
\pdfoutput=1
\usepackage{jheppub} 
\usepackage{multirow}
\usepackage{booktabs}
\usepackage{subcaption}

\title{Particle track reconstruction in high-density collider environments with an Evolving-Geometry Transformer}

\author[a,1]{Chenxiu Ou,\note{These authors contributed equally.}}
\author[b,c,1]{Yan Wang,}
\author[a]{Mingjiang Liang,}
\author[c]{Man-Hong Yung,}
\author[a,2]{Qin Zhang,\note{Corresponding author.}}
\author[d,2]{Zhaofeng Su}

\affiliation[a]{College of Computer Science and Software Engineering,\\
Shenzhen University, Shenzhen 518060, China}
\affiliation[b]{School of Computer Science and Technology,\\
University of Science and Technology of China, Hefei 230027, China}
\affiliation[c]{International Quantum Academy,\\
 Shenzhen 518048, China}
\affiliation[d]{Research Institute for Quantum Technology,\\
The Hong Kong Polytechnic University, Hong Kong, China}

\emailAdd{qinzhang@szu.edu.cn}
\emailAdd{youngpath2012@gmail.com}

\abstract{Charged particle track reconstruction at the High-Luminosity Large Hadron Collider is challenged by dense hit environments and combinatorial ambiguities. Existing graph-based and Transformer approaches commonly rely on static geometric neighbourhoods, which cannot adapt as track-level representations evolve. We propose the Evolving-Geometry Transformer (EGT), a graph Transformer that reconstructs the hit-connectivity graph at successive encoder layers and incorporates relative hit coordinates as learnable attention biases. Across five simulation benchmarks spanning linear, helical, and reduced TrackML events with up to 200--500 simultaneous trajectories, EGT achieves the highest FitAccuracy on three of five settings under the shared benchmark protocol, reaching 79.6\% on the most complex setting and exceeding the strongest baseline by 1.6 percentage points. Replacing the evolving topology with a static graph reduces the perfect-track rate from 81\% to 29\%, while removing the geometric bias reduces it to 73\%. Under synthetic background injection equal to the number of signal hits, FitAccuracy decreases by 3.4 percentage points. The neural-network forward pass requires 31~ms per event on an NVIDIA A100 GPU. These results indicate that iterative refinement of hit associations improves complete-track recovery when local geometry alone is insufficient to determine trajectory membership.}


\begin{document}
\maketitle
\flushbottom
\section{Introduction}\label{sec1}

Charged-particle track reconstruction is a central component of event reconstruction at collider experiments, where particle trajectories must be inferred from discrete spatial measurements recorded in layered tracking detectors~\cite{albrecht2019roadmap}. At the High-Luminosity Large Hadron Collider (HL-LHC), up to several hundred simultaneous proton--proton interactions per bunch crossing are expected~\cite{ApollinariG.:2017ojx}, producing dense hit environments and severe combinatorial ambiguity in trajectory formation. Together with stringent reconstruction latency constraints~\cite{atlas_hllhc_comp,cms_hllhc_comp}, these conditions motivate tracking methods that maintain high reconstruction efficiency while controlling computational growth with hit multiplicity.

Classical LHC tracking relies primarily on Kalman-filter techniques and their combinatorial extensions, including Combinatorial Kalman Filter (CKF) approaches used in ATLAS and CMS~\cite{billoir1984track,CMS-TRK-11-001,ATLAS:Run3Tracking2024}. These methods provide mature and well-characterised physics performance, but their computational cost can grow rapidly in dense events as multiple compatible hit associations are retained and propagated as track candidates. This combinatorial scaling has motivated machine-learning approaches that replace explicit hypothesis enumeration with learned representations of track structure.

Graph-based machine-learning methods provide a natural formulation of this problem by representing detector hits as graph nodes and candidate hit-to-hit associations as edges. Early GNN-based approaches demonstrated complete tracking pipelines involving graph construction, edge classification, and track reconstruction on the TrackML dataset~\cite{Ju2020GraphNN,trackml-particle-identification}. The Exa.TrkX programme subsequently developed scalable GNN-based pipelines targeting HL-LHC reconstruction workloads, demonstrating favourable scaling relative to traditional combinatorial approaches~\cite{Ju_2021,choma2020trackseedinglabellingembeddedspace}. Subsequent studies have explored hierarchical architectures and hardware-aware implementations for detector-specific applications such as the LHCb VELO system~\cite{Correia_2024}. A common feature of these approaches is that graph connectivity is initially determined from detector geometry, for example using $k$-nearest-neighbour ($k$NN) or fixed-radius constructions. More recent methods such as EggNet~\cite{calafiura2024eggnet} address the limitations of a fixed topology by dynamically updating the neighbourhood graph from learned node representations at successive layers. This establishes an important progression from static geometric graphs towards adaptive graph structures in which the set of interacting hit pairs can evolve during inference.

Transformer-based approaches provide an alternative means of modelling hit-to-hit interactions through self-attention over sets or sequences of detector measurements~\cite{melkani2024tracksortertransformerbasedsortingalgorithm,caron2025trackformers}. Full self-attention, however, scales as $\mathcal{O}(N^2)$ in the number of hits. Sparse attention schemes reduce this cost by restricting interactions through predefined detector-coordinate orderings, fixed windows, or related geometric constraints~\cite{md46-yqgd,haris2025knn}. Although effective computationally, such constraints impose a static notion of locality that may become restrictive when spatial proximity alone does not reliably indicate common trajectory membership, for example for strongly curved or closely spaced trajectories.

These developments expose two complementary questions in scalable hit-based tracking: \emph{which hit pairs should interact, and how strongly should they interact?} The first question concerns the interaction topology. Dynamic graph methods such as EggNet have demonstrated that the set of candidate interactions can be adapted from learned representations, moving beyond a fixed geometric graph. The second question concerns the strength of the interaction once a pair has been selected. In existing graph and sparse-attention tracking approaches, detector geometry is commonly used to define connectivity or locality, whereas explicit relative hit geometry is less commonly introduced as a dedicated learnable term that directly modulates pairwise attention weights. This distinction is important in dense tracking environments, where two hits may be close in detector coordinates without belonging to the same trajectory, while geometrically informative relative configurations can provide additional evidence for their relationship. Recent advances in Transformer architectures have shown that relative positional and structural information can instead be incorporated directly into attention through additive biases, allowing pairwise geometry to influence interaction strengths in a differentiable manner~\cite{shaw-etal-2018-self,ying2021transformersreallyperformbad,zhao2021point}. Such geometry-conditioned interaction mechanisms remain comparatively unexplored in charged-particle tracking.

The two questions are inherently coupled. Improved trajectory representations can provide a better basis for selecting relevant neighbours, while improved neighbourhoods provide the relational context needed to refine those representations. This circular dependence motivates an iterative formulation in which hit representations and their interaction structure evolve together.

In this work, we propose a graph Transformer architecture with evolving neighbourhoods for charged-particle track reconstruction. The model combines an adaptive graph structure, whose connectivity evolves with the learned representations, with an attention mechanism that directly incorporates relative detector geometry into pairwise interactions. The evolving topology allows the model to progressively refine which hit pairs exchange information, while geometry-aware attention provides a complementary mechanism for learning how strongly those interactions should contribute. This combination is designed to retain the scalability benefits of sparse graph-based processing while making more explicit use of the geometric structure of the detector during relational reasoning.

The main contributions of this work are threefold. First, we introduce an evolving-neighbourhood graph Transformer in which the interaction topology is progressively refined during encoding rather than being fixed by the initial geometric graph. Second, we introduce a geometry-aware attention mechanism that uses relative hit-pair coordinates as a learnable additive bias, allowing detector geometry to directly modulate interaction strength rather than serving only as a connectivity prior. Third, we demonstrate the effectiveness of the combined architecture through systematic experiments and ablations, showing that evolving neighbourhoods provide the primary improvement in resolving combinatorial ambiguities, while geometry-aware attention provides additional robustness for highly curved trajectories and densely populated detector regions. Together, these results demonstrate that adaptive topology and geometry-aware interaction provide complementary mechanisms for scalable track reconstruction in high-occupancy environments.

We evaluate EGT on five simulation benchmarks spanning linear and helical trajectories and increasing track multiplicity. Under the shared TrackFormers evaluation protocol, EGT achieves the highest FitAccuracy on three of five benchmarks, with the advantage becoming more pronounced in several geometrically ambiguous regimes. Most importantly, controlled ablations reveal a strong asymmetry between the two mechanisms: replacing the evolving topology with a static graph reduces the perfect-track rate from $0.81$ to $0.29$, whereas removing geometry-aware attention reduces it to $0.73$. These results indicate that iterative refinement of hit connectivity is the dominant contribution to complete-track recovery in the evaluated setting, with relative geometry providing a complementary signal.

\section{Detector environment}
\label{sec:detector}

Charged-particle trajectories in collider experiments are measured as spatially discrete hits in multilayer silicon tracking detectors operating inside a solenoidal magnetic field. The reconstruction problem is to associate these measurements across barrel and forward detector layers with their underlying particle trajectories.

In a nearly uniform magnetic field aligned with the beam axis, the transverse motion of a charged particle is approximately circular, while its motion along the beam direction remains approximately uniform. Neglecting energy loss and multiple scattering, the radius of curvature in the transverse plane is given by

\begin{equation}
R \simeq \frac{p_T}{0.3\,|q|B},
\label{eq:curvature}
\end{equation}
where $p_T$ is expressed in GeV/$c$, $B$ in tesla, $R$ in metres, and $q$ in units of the elementary charge. Lower-$p_T$ particles therefore exhibit stronger curvature, making simple spatial proximity a less reliable indicator of track membership, particularly in dense hit environments.

The transverse momentum is defined from the Cartesian momentum components as
\begin{equation}
p_T = \sqrt{p_x^2+p_y^2},
\label{eq:pt}
\end{equation}
while the particle direction relative to the beam axis is commonly characterised by the pseudorapidity
\begin{equation}
\eta = -\ln\left(\tan\frac{\theta}{2}\right),
\label{eq:eta}
\end{equation}
where $\theta$ is the polar angle with respect to the beam axis. The azimuthal angle is given by
\begin{equation}
\phi = \operatorname{atan2}(p_y,p_x).
\label{eq:phi}
\end{equation}
The quantities $p_T$ and $|\eta|$ provide useful measures of two important aspects of the tracking environment. The former is directly related to the curvature of charged-particle trajectories, while the latter characterises the direction of the particle relative to the beam axis and distinguishes central barrel from forward detector regions. Reconstruction performance is therefore evaluated as a function of both $p_T$ and $|\eta|$ throughout this work.

The benchmark datasets considered in this study provide complementary detector environments. REDVID~\cite{10.1007/978-3-031-63751-3_6} provides a configurable and simplified detector model that allows the geometric complexity of the tracking problem to be controlled systematically. In contrast, the TrackML benchmark~\cite{amrouche2020tracking,amrouche2023tracking} provides a substantially more realistic detector geometry and high-occupancy event environment representative of the challenges encountered in HL-LHC tracking. Together, these datasets allow the proposed architecture to be evaluated from controlled geometric configurations to realistic high-pile-up conditions.

\subsection{REDVID detector environment}
\label{sec:redvid}

The first group of benchmark datasets is generated using the REDVID simulation framework~\cite{10.1007/978-3-031-63751-3_6}, which provides a configurable detector model for controlled studies of charged-particle tracking. The detector geometry is defined in cylindrical coordinates $(r,\phi,z)$ and reproduces the main geometric characteristics of a multilayer silicon tracking detector while simplifying detector-specific engineering details.

\begin{figure}[t]
\centering
\includegraphics[width=0.7\linewidth]{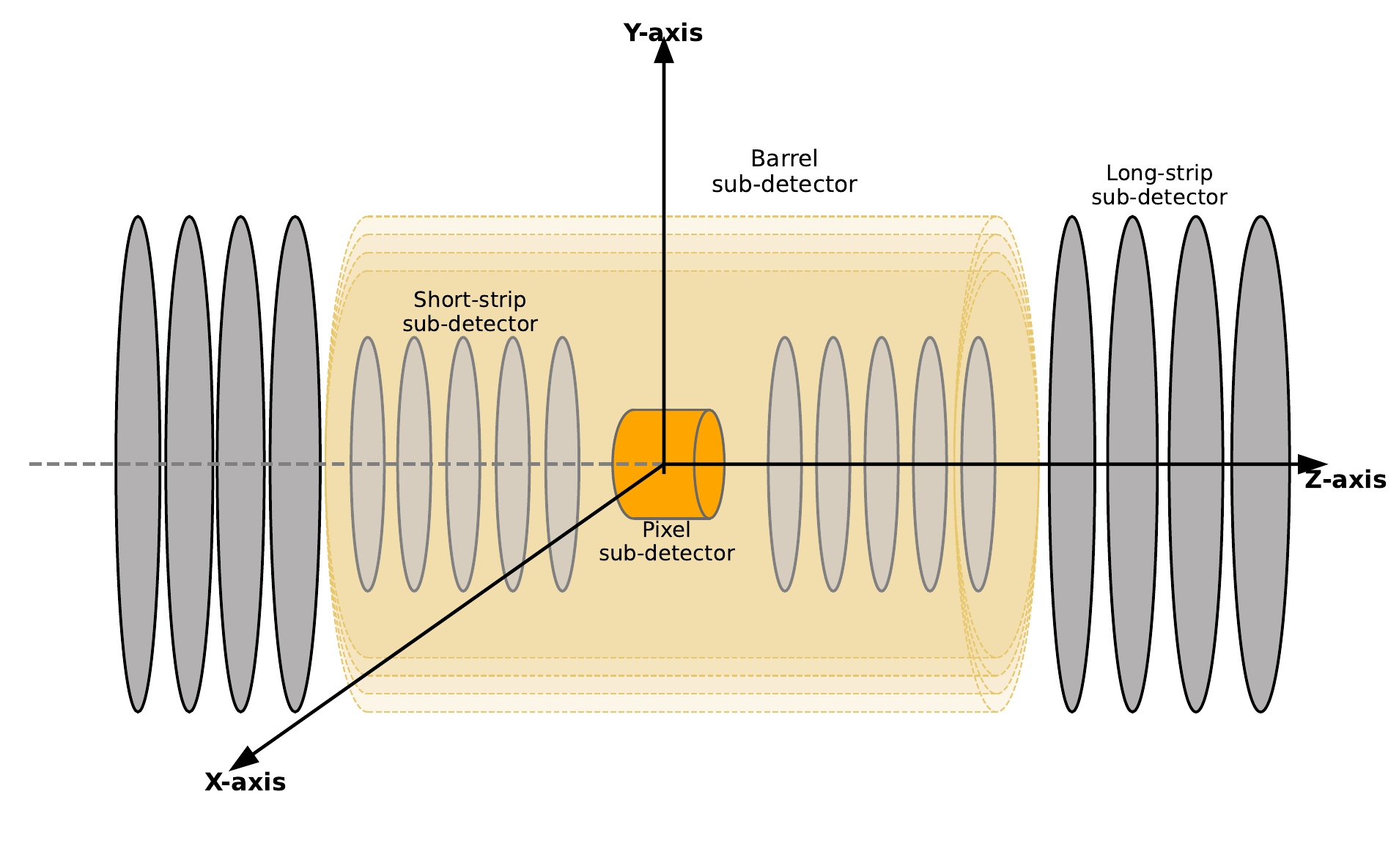}
\caption{
Schematic illustration of the tracker geometry used by the benchmark datasets, redrawn based on the detector configuration described in~\cite{caron2025trackformers}. The detector consists of concentric barrel layers and forward endcap disks surrounding the interaction region. Charged particles traverse the detector within a solenoidal magnetic field and produce discrete spatial measurements across multiple detector layers. The figure is intended for illustration and is not drawn to scale.
}
\label{fig:detector}
\end{figure}

The detector consists of concentric barrel layers together with forward disk detectors positioned symmetrically around the interaction region, as illustrated in figure~\ref{fig:detector}. Four detector element types are supported: Pixel, Barrel, Short-Strip, and Long-Strip detectors. Their dimensions and positions can be configured parametrically, allowing different detector geometries to be generated while retaining the characteristic layered structure relevant to track reconstruction.

Particle trajectories originate from the interaction region and are generated according to two propagation models. The first assumes straight-line trajectories and therefore provides an idealised reconstruction environment without magnetic-field-induced curvature. The second generates expanding helical trajectories that approximate the motion of charged particles in a solenoidal magnetic field. Gaussian smearing is subsequently applied to the generated hit coordinates to model finite detector resolution.

To isolate the geometric aspects of the reconstruction problem, the REDVID datasets used in this work assume a single proton--proton interaction per event and do not explicitly model pile-up interactions, multiple scattering, energy loss, or secondary interactions with detector material. These simplifications make REDVID a controlled test bed for isolating the effects of trajectory geometry and curvature on hit association.

\subsection{TrackML detector environment}
\label{sec:trackml}

To evaluate the proposed method in a more complex reconstruction environment,
we additionally use Two TrackML-derived benchmarks~\cite{amrouche2020tracking,amrouche2023tracking}, following the preprocessing protocol adopted in TrackFormers.

Compared with REDVID, these benchmarks provide a substantially more realistic silicon tracking geometry and higher track multiplicity, resulting in a more demanding hit-association problem.

The simulated detector consists of concentric cylindrical barrel layers surrounding the interaction region together with forward endcap disks, providing tracking coverage over a broad range of pseudorapidity. Charged particles propagate through the detector in a solenoidal magnetic field and produce discrete three-dimensional measurements as they traverse successive detector layers. In this work, detector hits are represented by their global Cartesian coordinates,
\begin{equation}
\mathbf{x} = (x,y,z),
\label{eq:hit_coordinates}
\end{equation}
which constitute the input features provided to the proposed model.

The TrackML-derived benchmarks contain substantially more trajectories per event than the REDVID datasets and incorporate a more complex detector geometry and particle-production environment. Consequently, spatially nearby hits are more likely to originate from different trajectories, increasing the ambiguity of hit association and providing a more stringent test of the learned relational structure.

TrackML additionally provides truth-level particle information, including the production vertex, momentum components $(p_x,p_y,p_z)$, electric charge $q$, and particle identity. These truth-level quantities are not provided to the model during inference. Instead, they are used to construct the training targets and to evaluate the reconstruction performance against the underlying simulated particle trajectories.

REDVID therefore provides controlled tests of the role of trajectory geometry and curvature, whereas the TrackML-derived benchmarks test the method under substantially greater track multiplicity and more realistic detector geometry. Together, they provide complementary regimes for evaluating whether adaptive relational reasoning becomes increasingly beneficial as hit-association ambiguity grows.

\section{Model architecture}
\begin{figure*}[t]
    \centering
    \makebox[\textwidth]{\includegraphics[width=1.0\textwidth]{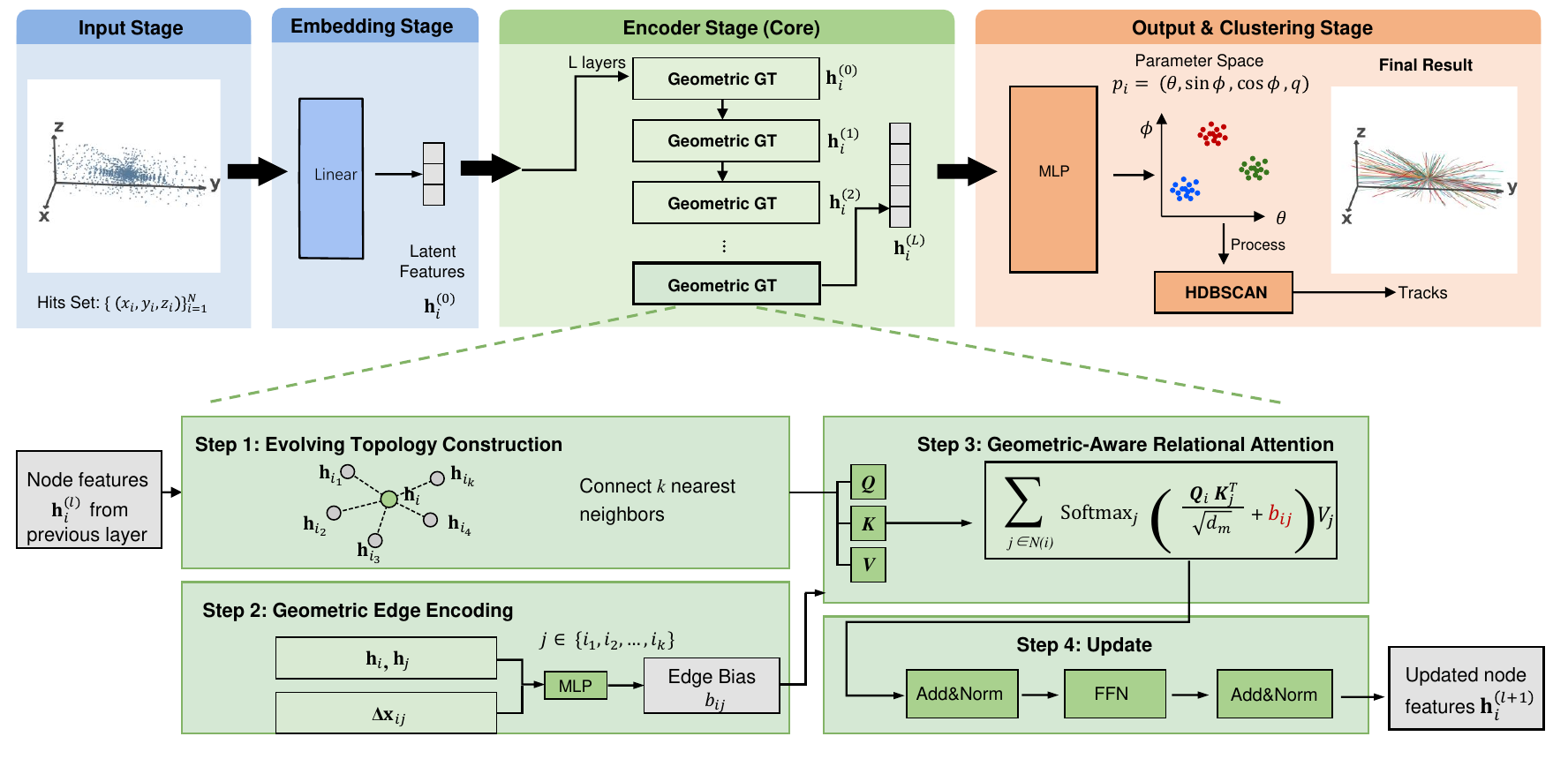}}
    \caption{Overview of the proposed Evolving-Geometry Transformer (EGT), with hit-level data shown in blue and learned operations in orange.
\textbf{(a)}$N$ input hits, each described by spatial coordinates $(x_i, y_i, z_i)$, are first projected into a latent space via a linear embedding layer.
\textbf{(b)}The encoder then applies $L$ stacked EGT blocks, each of which (i) rebuilds the interaction graph by connecting each hit to its $k$ nearest neighbours in the current latent space, allowing the relational topology to be recomputed from progressively updated hit representations; and (ii) updates hit representations through geometry-aware multi-head attention, in which relative coordinate offsets $\Delta\mathbf{x}_{ij}$ between hit pairs are encoded as additive biases on the attention logits, enabling the model to exploit local geometric compatibility as a continuous learnable signal.
\textbf{(c)}
After $L$ such layers, a linear output head maps the per-hit embeddings $\mathbf{h}_i^{(L)}$ to trajectory parameter predictions $(\theta, \sin\phi, \cos\phi, q)$, and HDBSCAN clustering groups the resulting predictions into particle track candidates.
A detailed view of a single EGT block is shown below.%
}
    \label{fig:architecture}
\end{figure*}

Figure~\ref{fig:architecture} illustrates the overall pipeline of EGT, which combines layer-wise topology refinement with geometry-aware feature interaction. Raw hit coordinates are first projected into a latent representation space via a linear  embedding layer. At each encoder layer, a $k$NN  graph is dynamically constructed from the current node representations, and geometry-aware relation features derived from relative coordinate offsets between hit pairs are injected as additive biases into multi-head attention. After $L$ encoder layers, a linear decoding head produces per-hit physical parameter predictions, which are subsequently grouped into particle tracks via HDBSCAN post-processing.

\subsection{Problem formulation}

Particle track reconstruction can be formulated as a structured regression problem over a set of detector measurements.  Given a collection of hits from a single collision event,
\begin{equation}
\mathcal{H} = \{ \mathbf{x}_i \}_{i=1}^N,
\end{equation}
each hit $x_i \in \mathbb{R}^3$ is associated with its spatial coordinates $(x_i, y_i, z_i)$ in the detector frame.

Our objective is to learn a function
\begin{equation}
F_\Theta : \mathbb{R}^{N\times 3} \rightarrow \mathbb{R}^{N\times 4}
\end{equation}
that predicts, for each hit $x_i$, the physical parameters of the particle trajectory to which it belongs. Specifically, the prediction for hit $i$ is a four-dimensional vector
\begin{equation}
\hat{\mathbf{p}}_i = F_\Theta(x_i),
\end{equation}
whose components are, respectively, the polar angle $\theta$, the azimuthal components $\sin\phi$ and $\cos\phi$, and the particle charge $q$, i.e.\ $\hat{\mathbf{p}}_i = (\theta,\,\sin\phi,\,\cos\phi,\,q)$. The predicted quantities $(\theta, \sin\phi, \cos\phi, q)$ provide a compact representation of the particle direction and charge. Specifically, $\theta$ determines the polar direction, $(\sin\phi, \cos\phi)$ encode the azimuthal direction, and $q$ specifies the particle charge. These quantities are used as the clustering coordinates for HDBSCAN to group hits into track candidates.

Although predictions are made at the hit level, the target quantities are shared by all hits originating from the same particle. Track reconstruction is therefore inherently relational: reliable prediction requires integrating information across multiple measurements associated with a common trajectory.

We model the event as an unknown latent graph
\begin{equation}
\mathcal{G}^* = (\mathcal{V}, \mathcal{E}^*),
\end{equation}
where each vertex $v_i \in \mathcal{V}$ corresponds to a detector hit $x_i \in \mathcal{H}$. An edge $(v_i, v_j) \in \mathcal{E}^*$ exists if hits $i$ and $j$ originate from the same physical particle. Since the true relational structure is not observed at inference time, the model must infer a topology that enables trajectory-consistent parameter regression.

This formulation highlights the central difficulty of particle tracking: trajectory-level information is distributed across spatially separated measurements, while the associations between those measurements are unknown a priori. Accurate reconstruction therefore requires the model to refine both the hit representations and the relational structure through which information is exchanged. This observation motivates the evolving graph mechanism introduced below.

\subsection{Evolving graph topology}

At each encoder layer $\ell$, we construct a layer-specific graph
\begin{equation}
\mathcal{G}^{(\ell)} = (\mathcal{V}, \mathcal{E}^{(\ell)}),
\end{equation}
where the vertex set $\mathcal{V}$ is fixed and the edge set $\mathcal{E}^{(\ell)}$ is induced by the current latent representations.

Let $\mathbf{h}_i^{(\ell)} \in \mathbb{R}^{d_h}$ denote the latent representation of hit $i$ at layer $\ell$. For each node $i$, we define a neighbourhood $\mathcal{N}_i^{(\ell)}$ by identifying its $k$NN in the latent feature space:
\begin{equation}
\mathcal{N}_i^{(\ell)} = \mathrm{kNN}\!\left(\mathbf{h}_i^{(\ell)}\right).
\end{equation}
The edge set at layer $\ell$ is then given by
\begin{equation}
\mathcal{E}^{(\ell)} = \{(i,j) \mid j \in \mathcal{N}_i^{(\ell)}\}.
\end{equation}

Under this formulation, the interaction topology is explicitly coupled to the learned representations. Rather than maintaining a fixed graph throughout the encoder, EGT reconstructs the neighbourhood structure at successive layers. The current graph determines how information is aggregated, while the updated representations in turn determine the graph used by the next layer.

This iterative process allows the relational structure to become progressively aligned with the emerging trajectory representation. Early-layer neighbourhoods are dominated by local similarities inherited from the input geometry, whereas deeper representations incorporate increasingly richer track-level context. Reconstructing the graph throughout the encoder therefore enables EGT to continually refine the relational context used for subsequent feature updates.

\subsection{Geometry-aware graph Transformer}

The evolving graph topology determines which hit pairs exchange information, but not how strongly individual neighbours should contribute. We therefore complement the adaptive topology with geometry-aware attention. Given the layer-specific graph $\mathcal{G}^{(\ell)}$, attention for hit $i$ is restricted to $\mathcal{N}_i^{(\ell)}$, while relative hit geometry provides an additional signal for modulating pairwise interactions.

For each edge $(i,j) \in \mathcal{E}^{(\ell)}$, we construct a geometry-aware relation feature
\begin{equation}
\mathbf{e}_{ij}^{(\ell)} = \psi\!\left(\mathbf{h}_i^{(\ell)}, \mathbf{h}_j^{(\ell)}, \Delta\mathbf{x}_{ij}\right),
\end{equation}

where $\Delta\mathbf{x}_{ij}=\mathbf{x}_i-\mathbf{x}_j$ denotes the relative coordinate offset between hits $i$ and $j$, and $\psi$ is an MLP producing a $d_e$-dimensional relation embedding. The resulting representation combines latent node information with pairwise geometry to evaluate local hit compatibility.

For each attention head $m \in \{1, \ldots, M\}$, with per-head dimension $d_m$, the query, key, and value vectors are computed as
\begin{align}
\mathbf{q}_i^m &= W_Q^m \mathbf{h}_i^{(\ell)}, \\
\mathbf{k}_j^m &= W_K^m \mathbf{h}_j^{(\ell)}, \\
\mathbf{v}_j^m &= W_V^m \mathbf{h}_j^{(\ell)}.
\end{align}

Here, $i$ indexes the query hit and $j$ ranges over all nodes; neighbourhood constraints on $j$ are imposed in the subsequent attention computation over $\mathcal{N}_i^{(\ell)}$. We then incorporate the geometry-aware relation feature into the attention logits through an additive bias:
\begin{equation}
\alpha_{ij}^m =
\frac{(\mathbf{q}_i^m)^\top \mathbf{k}_j^m}{\sqrt{d_m}}
+\lambda \cdot b^m\!\left(\mathbf{e}_{ij}^{(\ell)}\right),
\end{equation}
where $b^m : \mathbb{R}^{d_e} \to \mathbb{R}$ maps the relation feature to a scalar bias. In this way, attention is conditioned not only on latent node similarity, but also on the geometric relationship between hits.

The normalized attention weights are computed over the local neighbourhood:
\begin{equation}
a_{ij}^m =
\frac{\exp(\alpha_{ij}^m)}
{\sum\limits_{j' \in \mathcal{N}_i^{(\ell)}} \exp(\alpha_{ij'}^m)}.
\end{equation}

The multi-head aggregated message is then given by
\begin{equation}
\mathbf{u}_i^{(\ell)} =
W_O \,
\mathrm{Concat}_{m=1}^M
\left(
\sum_{j \in \mathcal{N}_i^{(\ell)}} a_{ij}^m \mathbf{v}_j^m
\right),
\end{equation}
where $W_O \in \mathbb{R}^{d_h \times (M d_m)}$ is the output projection matrix. The multi-head attention output is combined with the input representation through a residual connection, followed by layer normalization and a position-wise feed-forward network, as in standard Transformer architectures, to produce the next-layer representation $\mathbf{h}_i^{(\ell+1)}$.

The explicit use of relative geometry is particularly beneficial in tracking, where local displacement patterns may reflect directional consistency and curvature-related structure along particle trajectories.In the solenoidal magnetic field of strength $B$, a charged particle follows a helical trajectory with transverse Larmor radius $r = p_T / (0.3|q|B)$. The azimuthal deflection between hits on adjacent detector layers separated by radial distance $d$ is approximately $d/r \propto p_T^{-1}$, so the sequence of relative displacement vectors $\Delta\mathbf{x}_{ij}$ encodes curvature information directly linked to $p_T$; the geometry-aware bias therefore provides the model with an implicit $p_T$-sensitive prior that is physically grounded in the Lorentz force acting on the charged particle.

\section{Dataset and experimental setup}

\subsection{Datasets and baselines}

We evaluate EGT on the five benchmark datasets introduced in TrackFormers~\cite{caron2025trackformers}. The first three datasets are generated with REDVID and contain 10--50 linear tracks, 10--50 helical tracks, and 50--100 helical tracks per event, respectively. The remaining two datasets are reduced from TrackML and contain 10--50 and 200--500 tracks per event. Together, these datasets provide a controlled progression in trajectory complexity and track multiplicity.

For each dataset, we follow the event construction, preprocessing procedure, hit-level representation, supervision targets, and train/validation/test split adopted in TrackFormers. We further evaluate our approach against the five baseline models from \cite{caron2025trackformers}, namely EncDec, EncCla, EncReg, EncReg-FA, and U\text{-}Net. Although these methods differ in their preprocessing pipelines and supervision targets, they are all evaluated on the same underlying datasets under a common evaluation metric. As such, they provide a standardized benchmark for comparing different modeling approaches on this task.The implementation reuses the data-processing, training, and evaluation framework of the publicly available TrackFormers codebase~\cite{caron2025trackformers}. The original Transformer regressor is replaced by the proposed EGT architecture, including the evolving $k$-NN graph construction and geometry-aware attention mechanism.

We note that several concurrent works—including EggNet \cite{calafiura2024eggnet} and MaskFormer-based approaches—have also been applied to particle track reconstruction, in some cases achieving strong performance on TrackML-derived datasets. However, as observed in TrackFormers \cite{caron2025trackformers}, the lack of a unified data reduction and preprocessing protocol across the community makes direct numerical comparison between methods trained under different pipelines unreliable. Our experimental setup strictly follows the protocol established in TrackFormers, using identical event construction, hit-level representations, supervision targets, and dataset splits. Comparisons are therefore restricted to models evaluated under this shared protocol; cross-protocol comparisons are left to future work pending standardization of community benchmarks. Dataset-specific truth information is used to construct the hit-level regression targets following the preprocessing protocol of TrackFormers~\cite{caron2025trackformers}.

A differential breakdown of the per-particle FitAccuracy score as a function of $|\eta|$ and $p_T$ is provided in section~\ref{sec:diffperf}.

\subsection{Training protocol}
Full training configurations, including dataset-specific learning rates, batch sizes, and scheduling strategies, are provided in appendix~\ref{app:impl}.

\subsection{Evaluation and post-processing}
For reconstruction-level evaluation, we adopt the FitAccuracy score following the protocol of TrackFormers~\cite{caron2025trackformers}. A reconstructed track is matched to a true particle if more than half of its hits originate from that particle and more than half of that particle's hits are contained within the track; FitAccuracy is the sum of hit weights $w_i$ over all matched hits, normalised by the total truth hit weight in the event. For TrackML-derived datasets, $w_i$ is the per-hit physics weight provided in the TrackML truth tables; for REDVID-derived datasets, which carry no such weights, $w_i = 1$ is used for all hits.

The ablation tables in section~\ref{sec:results} also report three auxiliary metrics. \emph{Perfects} is the fraction of true particles whose hits are entirely recovered by a single reconstructed track with less than 1\% contamination. \emph{Doubles} is the fraction of true particles satisfying the bidirectional majority criterion above, i.e.\ the standard per-particle reconstruction efficiency. \emph{LHCs} is the fraction of reconstructed tracks in which more than 75\% of hits originate from the same true particle, with reconstructed tracks rather than true particles as the denominator. We compare EGT against the five published baseline results reported in TrackFormers~\cite{caron2025trackformers} whenever available.

For post-processing, we apply HDBSCAN clustering to the predicted hit-level outputs. Since the exact post-processing hyperparameters used by EncReg~\cite{caron2025trackformers} are not reported in the paper and are not available in the released code, we independently tune \texttt{min\_cluster\_size} and \texttt{min\_samples} for our model on the validation set.

A systematic sweep over \texttt{min\_cluster\_size} and \texttt{min\_samples} on the TrackML 10--50 validation set is presented in figure~\ref{fig:hdbscan_sweep}; a detailed physical interpretation of the observed sensitivity is given in appendix~\ref{app:hdbscan}. The selected configuration (\texttt{min\_cluster\_size}$\,=5$, \texttt{min\_samples}$\,=3$) achieves FitAccuracy~$=0.9384$ with a fake rate of $2.3\%$ and is used for all reported results on this dataset.

\section{Results}\label{sec:results}

\subsection{Comparison on FitAccuracy}

Table~\ref{tab:accuracy_comparison} reports FitAccuracy on the five benchmark datasets. Under the shared evaluation protocol of TrackFormers \cite{caron2025trackformers}, EGT achieves the highest FitAccuracy on three of the five benchmarks, matches the best reported score on the linear REDVID setting, and remains competitive on TrackML 10--50.

The smaller gain on linear tracks is consistent with the lower
association ambiguity of this setting, where geometric proximity
already provides a strong cue for track membership.

The advantage becomes larger for curved and higher-multiplicity benchmarks, where geometrically nearby hits need not belong to the same trajectory. This trend is consistent with the intended role of the evolving topology and geometry-aware interaction mechanism, which are designed to refine hit associations as trajectory-level representations emerge.

\begin{table}[t]
\centering

\renewcommand{\arraystretch}{1.0}
\setlength{\tabcolsep}{3.4pt}
\begin{tabular}{lcccccc}
\toprule
\multirow{2}{*}{Dataset} & \multicolumn{5}{c}{Baseline Models (\%)} & \multirow{2}{*}{Ours (\%)} \\
\cmidrule(lr){2-6}
& EncDec & EncCla & EncReg & EncReg-FA & U\text{-}Net & \\
\midrule
REDVID -- 10--50 linear  & 93 & 93 & \textbf{97} & -- & 68 & \textbf{97.0} \\
REDVID -- 10--50 helical  & 85 & 93 & 92 & -- & 62 & \textbf{95.1} \\
REDVID -- 50--100 helical & 85 & 88 & 85 & -- & 57 & \textbf{88.3} \\
TrackML -- 10--50      & 26 & \textbf{94} & 93 & -- & -- & 93.8 \\
TrackML -- 200--500     & -- & 78 & 70 & 67 & -- & \textbf{79.6} \\
\bottomrule
\end{tabular}

\caption{FitAccuracy score comparison across the five benchmark datasets under the evaluation protocol of TrackFormers \cite{caron2025trackformers}. Best results among models evaluated under this protocol are shown in bold. Baseline results are \cite{caron2025trackformers}.}
\label{tab:accuracy_comparison}
\end{table}

\subsection{Differential performance analysis}\label{sec:diffperf}

\begin{figure*}[t]
    \centering
    \begin{subfigure}[b]{0.49\textwidth}
        \centering
        \includegraphics[width=\linewidth]{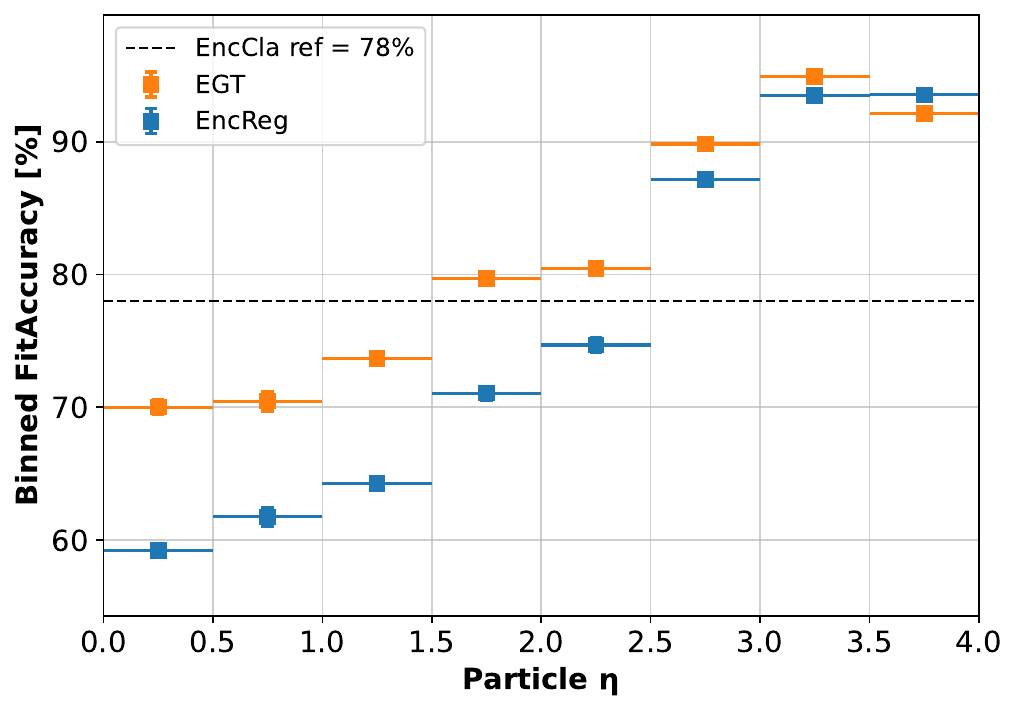}
        \caption{FitAccuracy vs.\ pseudorapidity $|\eta|$.}
        \label{fig:score_vs_eta}
    \end{subfigure}
    \hfill
    \begin{subfigure}[b]{0.49\textwidth}
        \centering
        \includegraphics[width=\linewidth]{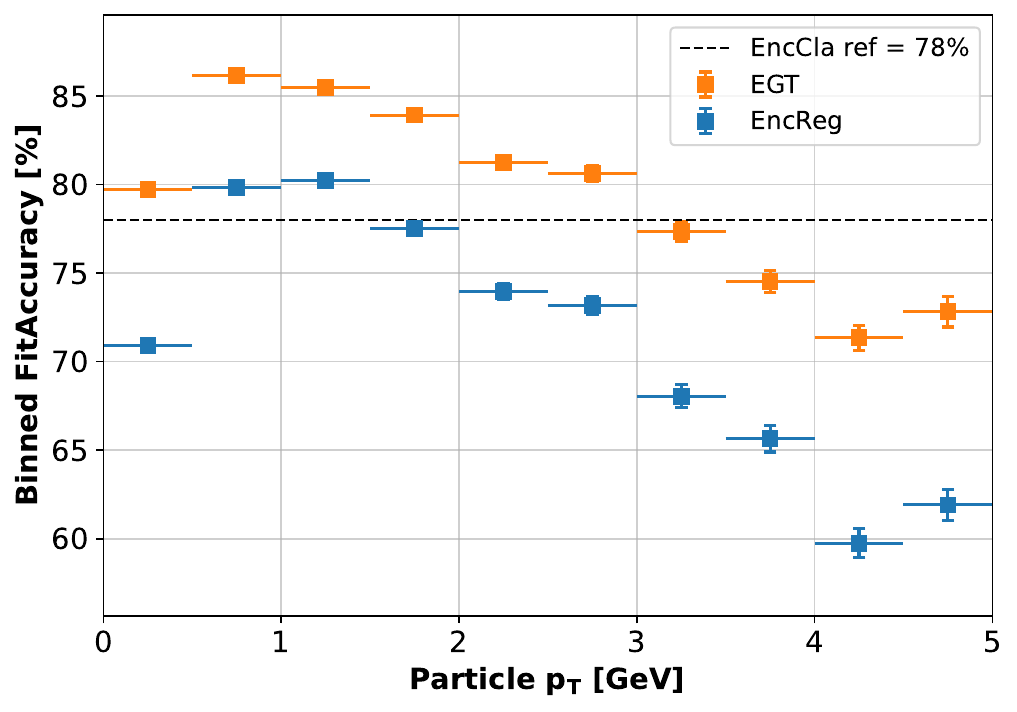}
        \caption{FitAccuracy vs.\ transverse momentum $p_T$.}
        \label{fig:score_vs_pt}
    \end{subfigure}
    \caption{Mean per-particle FitAccuracy score on the TrackML
    200--500 benchmark as a function of (a) particle pseudorapidity
    $|\eta|$ and (b) transverse momentum $p_T$. Orange and blue
    markers denote EGT and EncReg, respectively; horizontal bars
    span the bin width and vertical error bars indicate 95\%
    confidence intervals on the bin mean. Only particles satisfying
    the good-track selection criterion are included. In (b), the
    final bin includes overflow. The horizontal dashed line at 78\%
    marks the event-level FitAccuracy of EncCla, shown for
    reference.}
    \label{fig:score_vs_eta_pt}
\end{figure*}

Figure~\ref{fig:score_vs_eta_pt} shows the differential FitAccuracy of EGT and EncReg as functions of particle pseudorapidity and transverse momentum on the TrackML 200--500 benchmark.

\subsubsection{Performance vs.\ pseudorapidity}

Both models improve monotonically with $|\eta|$ up to $|\eta|\approx 3.25$. In the central barrel ($|\eta|\lesssim 0.5$), charged particle trajectories traverse the detector nearly perpendicular to the beam axis, maximising the number of competing pile-up tracks per unit solid angle. The resulting combinatorial background is most severe here: tracks from different primary vertices share the same local detector volume, producing the highest rate of ambiguous hit assignments across the full acceptance. EGT achieves 70.0\% in this region against 59.2\% for EncReg (${\approx}11$ percentage point gain).

The larger performance gap in the central region suggests that iterative relational refinement becomes increasingly beneficial as hit-association ambiguity grows. When hits from different trajectories are spatially interleaved, geometric proximity alone provides an incomplete cue for determining track membership. EGT addresses this ambiguity by updating the relational context across encoder layers, allowing hit representations and their neighbourhood structure to be progressively refined. This iterative process is conceptually analogous to successive track-state updates in classical tracking, although the refinement in EGT occurs in a learned latent representation.

Toward larger $|\eta|$, the performance gap decreases, and the two models converge to approximately 93--94\%. In the most forward bin ($|\eta|\in[3.5,4.0]$), their scores differ by only about one percentage point. The reduced gain is consistent with a regime in which local geometry provides a stronger cue for track association.

\subsubsection{Performance vs.\ transverse momentum}
Both models exhibit a non-monotonic profile, peaking near $p_T\approx 0.75$--$1.25$~GeV, reflecting increased reconstruction difficulty at opposite ends of the momentum spectrum.

At low $p_T$ below approximately $0.5$~GeV, the Larmor radius $r = p_T/(0.3|q|B) \approx 1.67\,p_T$~m in the 2~T solenoidal field becomes small. Strongly curved trajectories may traverse fewer detector layers, reducing the number of hits available for reliable hit association. Multiple scattering and energy loss may further perturb the trajectory. These effects provide a plausible explanation for the lower reconstruction performance observed in this region.

The relative displacement between hits contains curvature-related information that is correlated with $p_T$. EGT incorporates such pairwise geometric information explicitly through its geometry-aware attention mechanism, whereas EncReg operates on hit coordinates without an explicit relative-geometry bias. The larger performance gap in this region is consistent with a benefit from this additional geometric information. The score gap at $p_T<0.5$~GeV is $8.4$ percentage points.

At high $p_T$ above approximately $2$~GeV, the Larmor radius exceeds 3.3~m and the trajectories become progressively less curved. Both models show reduced FitAccuracy in this regime, while the performance gap in favour of EGT increases. As the curvature becomes weaker, local geometric differences between candidate trajectories may provide less distinctive information for hit association. The observed trend is consistent with a greater benefit from iterative neighbourhood refinement in this regime. The performance gap reaches approximately 11 percentage points above 4~GeV.

\subsection{Ablation studies}

The ablation results in Table~\ref{tab:ablation_trackml} reveal a pronounced asymmetry between the two core components of EGT. Removing the geometry-aware bias reduces FitAccuracy from $0.9388$ to $0.9029$ and Perfects from $0.81$ to $0.73$, whereas replacing the evolving topology with a static graph reduces FitAccuracy to $0.7864$ and Perfects to $0.29$. The substantially larger degradation caused by a static topology indicates that adaptive neighbourhood refinement is the dominant contribution to complete-track recovery in this setting, while geometry-aware attention provides a complementary improvement.

Removing only the geometric bias (w/o GB) produces a uniform degradation across all four metrics: FitAccuracy falls to $0.9029$, Perfects to $0.73$, Doubles to $0.89$, and LHC purity to $0.95$.

The more moderate degradation observed without the geometry-aware bias  suggests that relative hit geometry acts primarily as a complementary discriminative cue, while the evolving topology retains most of the structural benefit of the full model.

Replacing the evolving graph with a static topology (w/o EG) causes a qualitatively different and far more severe degradation, even when the geometric bias is retained: FitAccuracy drops to $0.7864$ and Perfects collapses to $0.29$, while LHC purity falls to $0.91$ and Doubles to $0.87$. The asymmetry between the Perfects and Doubles metrics is physically informative: Doubles measures only whether a particle is matched by any reconstructed track, whereas Perfects additionally requires that all of the particle's hits are recovered by a single cluster with less than $1\%$ contamination.

The disproportionate reduction in Perfects relative to Doubles is consistent with increased track fragmentation and contamination when the interaction topology is held fixed. This behaviour indicates that a static neighbourhood is substantially less effective at preserving complete trajectory-level associations than the evolving topology.

Removing both components (w/o EG \& GB) reduces FitAccuracy further to $0.7361$ with Perfects at $0.25$, confirming that geometry-aware interactions provide additional benefit when the topology is static.

\begin{table}[t]
\centering
\small
\setlength{\tabcolsep}{3pt}
\renewcommand{\arraystretch}{1.05}

\begin{tabular}{lcccccc}
\toprule
Variant & EG & GB & FitAccuracy & Perfects & Doubles & LHCs \\
\midrule
Full model   & \checkmark & \checkmark & 0.9388 & 0.81 & 0.91 & 0.97 \\
w/o GB       & \checkmark & \texttimes & 0.9029 & 0.73 & 0.89 & 0.95 \\
w/o EG       & \texttimes & \checkmark & 0.7864 & 0.29 & 0.87 & 0.91 \\
w/o EG \& GB & \texttimes & \texttimes & 0.7361 & 0.25 & 0.84 & 0.86 \\
\bottomrule
\end{tabular}

\caption{Ablation study on the TrackML 10--50 dataset. EG denotes evolving graph topology and GB denotes geometric bias.}
\label{tab:ablation_trackml}

\end{table}

\subsubsection{Effect of graph update frequency}

Table~\ref{tab:gstep_analysis} compares a static topology with graph updates every one, two, and three encoder layers.

The results show that dynamic topology refinement is essential for EGT. 

When graph updates are disabled, FitAccuracy drops sharply from above $0.9388$ to $0.7864$ and the Perfects metric decreases from $0.81$ to $0.29$. This indicates that a static interaction graph is insufficient for capturing the evolving relational structure required for accurate track reconstruction.

\begin{table}[t]
\centering

\small
\setlength{\tabcolsep}{4pt}
\begin{tabular}{lcccc}
\toprule
Update Policy & FitAccuracy & Perfects & Doubles & LHCs \\
\midrule
Every layer     & \textbf{0.9388} & \textbf{0.81} & \textbf{0.91} & \textbf{0.97} \\
Every 2 layers  & 0.9350          & 0.80          & \textbf{0.91} & \textbf{0.97} \\
Every 3 layers  & 0.9282          & 0.78          & \textbf{0.91} & 0.96          \\
\midrule
Static + GB             & 0.7864          & 0.29          & 0.87          & 0.91          \\
\bottomrule
\end{tabular}

\caption{Parameter analysis of graph update frequency on TrackML 10--50. Best results are highlighted in bold.}
\label{tab:gstep_analysis}

\end{table}

Among the dynamic variants, updating the graph at every layer achieves the best overall performance, with the highest FitAccuracy and Perfects. Updating every two layers remains highly competitive, suggesting that the main benefit comes from enabling iterative topology refinement rather than strictly requiring an update at every layer. In contrast, less frequent updates lead to a more noticeable performance drop, indicating that stale neighbourhood structures become less aligned with the current latent representation.

\subsubsection{Effect of geometry-aware attention bias}

We next study the effect of the geometry-aware attention bias coefficient $\lambda$, which scales the additive relation-based bias in the attention logits. Table~\ref{tab:ablation_edgebias} summarizes the results.

\begin{table}[t]
\centering

\renewcommand{\arraystretch}{1.15}
\setlength{\tabcolsep}{4pt}
\begin{tabular}{ccccc}
\toprule
$\lambda$ &  FitAccuracy & Perfects & Doubles & LHCs \\
\midrule
0.4 &  0.9336 & 0.80 & 0.91 & 0.97 \\
0.8 &  0.9371 & 0.81 & 0.91 & 0.97 \\
1.2 &  \textbf{0.9408} & \textbf{0.81} & \textbf{0.92} & \textbf{0.97} \\
1.6 & 0.9382 & \textbf{0.81} & 0.91 & \textbf{0.97} \\
2.0 & 0.9349 & \textbf{0.81} & 0.91 & \textbf{0.97} \\
\bottomrule
\end{tabular}%

\caption{Ablation on the geometry-aware attention bias coefficient $\lambda$ on TrackML 10--50 tracks. The coefficient $\lambda$ scales the additive geometry-aware bias in the attention logits.}
\label{tab:ablation_edgebias}

\end{table}

Removing the geometry-aware bias entirely ($\lambda = 0$) leads to a clear degradation in all metrics, with FitAccuracy falling from $0.9388$ to $0.9029$ and Perfects dropping from $0.81$ to $0.73$. This confirms that explicitly injecting geometric relations into attention is an important component of the proposed architecture.

Performance varies only modestly over $\lambda=0.4$--$2.0$, with the highest FitAccuracy obtained at $\lambda=1.2$. In contrast, removing the geometry-aware bias entirely produces a clear degradation, reducing FitAccuracy from $0.9388$ to $0.9029$. These results indicate that relative geometry provides a useful complementary signal, while the precise bias strength is not a dominant sensitivity of the model.

Overall, these ablations support the central design principle of our method: graph topology should co-evolve with node representations, and geometry-aware relations should act as a complementary bias that guides, rather than dominates, content-based attention.

\subsection{Robustness to synthetic background hits}
\label{sec:noise_robustness}

Throughout this work, the input to the model consists exclusively of truth-matched hits, with no hits lacking a truth-particle association present in the processed dataset. To probe performance under increasing detector occupancy, we perform a controlled stress test on the held-out test split of the TrackML 200--500 tracks dataset, comparing EGT against the EncCla hit-classification baseline~\cite{caron2025trackformers} under an identical noise protocol.

For each event, we inject additional background hits by uniform random sampling inside the axis-aligned bounding box of the signal hits, with a 10\% spatial margin, at fractions of 0\%, 10\%, 20\%, 50\% and 100\% relative to the number of signal hits per event. This injection scheme is an intentional simplification; genuine pile-up deposits follow the detector acceptance and particle-production angular distributions, whereas the injected hits here are drawn from a uniform spatial prior. Injected hits carry no truth label and zero reconstruction weight; they are presented to both models as ordinary inputs but are excluded from the FitAccuracy denominator.

\begin{figure}[t]
  \centering
  \includegraphics[width=0.62\linewidth]{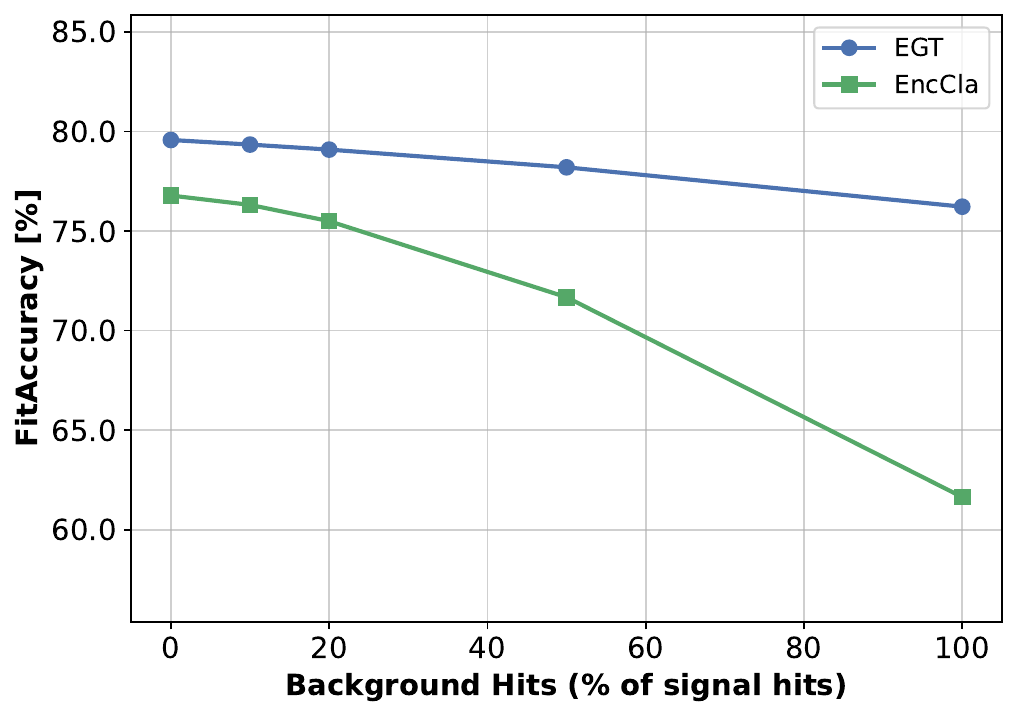}
  \caption{FitAccuracy on the TrackML 200--500 test set as a function of synthetically injected background hits, expressed as a fraction of signal hits per event. Background hits are sampled uniformly within the bounding box of the signal hits and carry no truth label. Blue markers denote EGT; green markers denote the EncCla baseline.
  }
  \label{fig:noise_robustness}
\end{figure}

Figure~\ref{fig:noise_robustness} shows that EGT degrades substantially more slowly than EncCla as the background fraction increases. At zero injected noise, EGT reaches a FitAccuracy of 79.6\%, compared with 76.8\% for EncCla. Under 50\% background injection, EGT retains 78.2\% FitAccuracy, a decrease of 1.4 percentage points; EncCla falls to 71.7\%, a decrease of 5.1 percentage points. At 100\% injection, where background hits equal signal hits in number, EGT still achieves 76.2\% FitAccuracy, a decrease of 3.4 percentage points from the unperturbed setting, while EncCla drops to 61.6\%, a decrease of 15.2 percentage points.

This difference is consistent with the distinct track-formation mechanisms used by the two models. EncCla produces a track-class prediction for every input hit, so the injected background hits are processed together with the signal hits and must also be assigned to the predefined track-class space. Because final tracks are obtained by density-based clustering in the predicted $(\theta,\sin\phi,\cos\phi,q)$ space, background hits would need to form trajectory-consistent predictions to contribute coherently to a reconstructed cluster. This provides a plausible explanation for the smaller degradation observed for EGT. The comparatively shallow degradation of EGT shows that its reconstruction performance is less sensitive to this form of synthetic background contamination than that of EncCla. This result provides an initial indication of robustness to increased hit occupancy, although the synthetic injection used here does not reproduce the spatial structure of realistic HL-LHC pile-up.

\subsection{Inference latency and computational cost}
\label{sec:inference}

Instance-level inference latency is measured following the protocol of TrackFormers~\cite{caron2025trackformers}: each test event is processed independently (batch size~$=1$), the first event is discarded as warmup, and means are taken over the held-out test split (15\%, seed~37). GPU time covers the neural-network forward pass only; CPU time includes HDBSCAN track clustering. EncReg latencies are taken from TrackFormers~\cite{caron2025trackformers} (NVIDIA A100 40\,GB, 18 CPU cores); all EGT numbers are measured in this work on an NVIDIA A100 80\,GB PCIe GPU. To enable a fair comparison, an inference-only implementation replaces the dense attention mask with sparse $k$-NN aggregation, restricting attention to $k$ neighbours per query and operating only on non-padded hits; reconstruction performance is unchanged relative to the training-time forward pass, with mean event-level FitAccuracy differing by less than $10^{-4}$ across all five datasets. Full implementation details are provided in appendix~\ref{app:impl}.

Figure~\ref{fig:inference_time} shows mean per-event GPU and CPU wall-clock time for both models across all five benchmark datasets.

\begin{figure*}[t]
    \centering
    \begin{subfigure}[b]{0.49\textwidth}
        \centering
        \includegraphics[width=\linewidth]{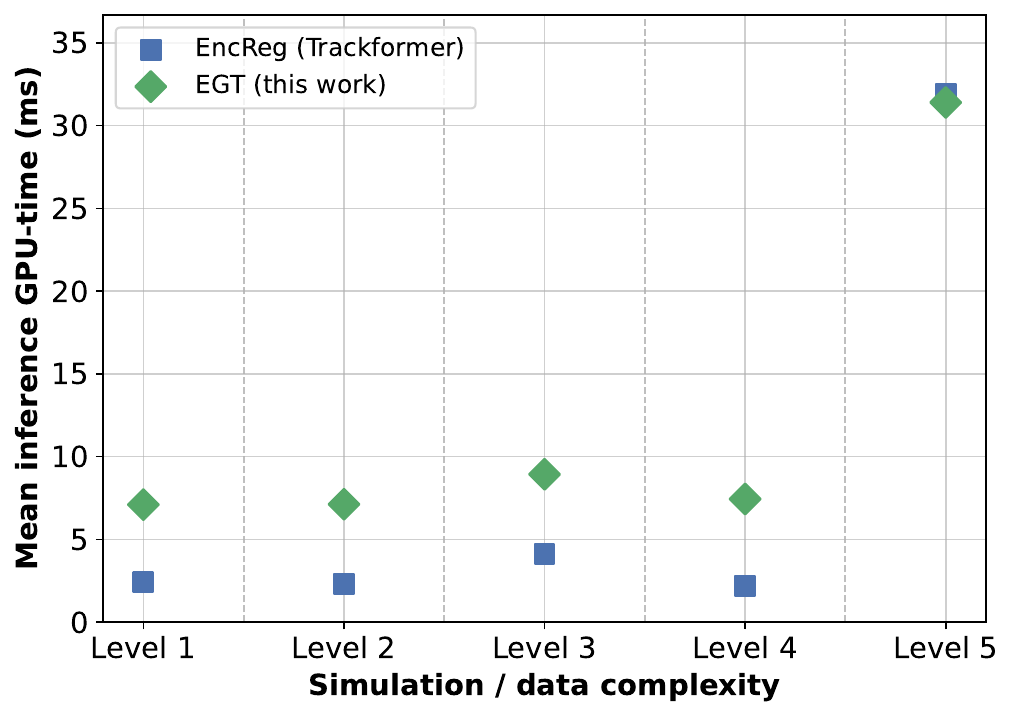}
        \caption{Mean per-event GPU inference time (ms).}
        \label{fig:inference_gpu}
    \end{subfigure}
    \hfill
    \begin{subfigure}[b]{0.49\textwidth}
        \centering
        \includegraphics[width=\linewidth]{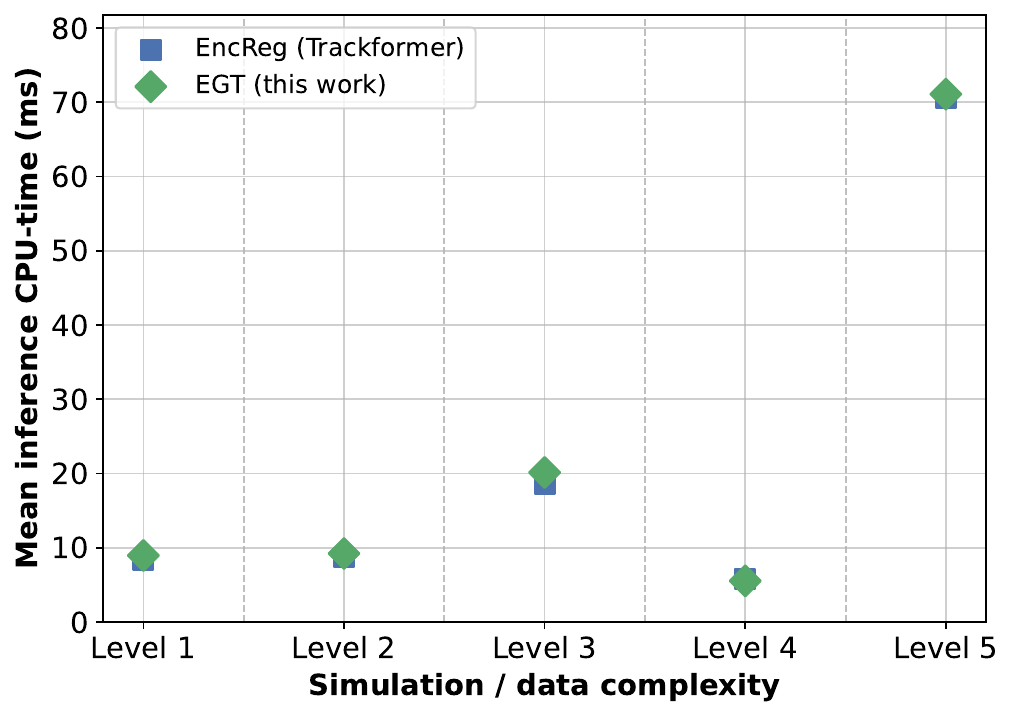}
        \caption{Mean per-event CPU time (ms).}
        \label{fig:inference_cpu}
    \end{subfigure}
    \caption{Per-event inference latency of EGT (this work, green diamonds) and EncReg~\cite{caron2025trackformers} (blue squares) across the five benchmark datasets, ordered by complexity. Levels 1--5 correspond to: REDVID 10--50 linear, REDVID 10--50 helical, REDVID 50--100 helical, TrackML 10--50, and TrackML 200--500. EncReg values are taken from TrackFormers~\cite{caron2025trackformers} (NVIDIA A100 40\,GB, 18 CPU cores); EGT values are measured in this work (NVIDIA A100 80\,GB PCIe, full test split, batch size $=1$). GPU time covers the neural-network forward pass only; CPU time includes HDBSCAN track clustering.}
    \label{fig:inference_time}
\end{figure*}

On GPU, EGT incurs a 2.2--3.4$\times$ overhead relative to EncReg on low-multiplicity datasets, attributable to per-layer $k$-NN graph reconstruction and edge-MLP evaluation. This overhead vanishes at the TrackML 200--500 scale, where EGT and EncReg show comparable GPU latency at 31.40\,ms and 31.9\,ms per event, respectively. At this scale, both models remain below 100\,ms per event on the hardware considered here.

On CPU, EGT and EncReg show similar latency across the benchmark datasets. The reported CPU component primarily reflects host-side post-processing, including hit unpadding and HDBSCAN clustering, whereas graph construction and attention are included in the GPU forward pass. Profiling at the highest complexity setting (TrackML 200--500, $k{=}64$, seven encoder layers, ${\sim}1\,695$ valid hits per event) identifies the edge MLP as the primary optimisation target, accounting for approximately 64\% of forward-pass GPU time; the remainder is distributed over sparse attention, per-layer $k$-NN reconstruction, and feed-forward blocks.

\section{Discussion}

\subsection{Scalability and the path to HL-LHC deployment}
\label{sec:anchor}

The inference results in section~\ref{sec:inference} are obtained on benchmarks where the hardest setting contains up to 500 simultaneous trajectories and approximately 1\,695 valid detector hits per event after preprocessing. Under full HL-LHC operating conditions a single bunch crossing is expected to produce on the order of $10^5$ detector hits, representing roughly a 60-fold increase in event occupancy relative to the largest benchmark studied here. The present results therefore do not yet constitute a demonstration of scalability to full HL-LHC event sizes.

The current implementation contains one $\mathcal{O}(N^2)$ operation. Exact $k$-NN graph reconstruction is performed at each encoder layer and requires computing all pairwise distances among $N$ hits. Once the graph is available, both the sparse attention computation and the edge MLP evaluate only the $Nk$ edges in $\mathcal{E}^{(\ell)}$ and therefore scale as $\mathcal{O}(Nk)$ with $k \ll N$; these stages do not require modification for deployment at larger scale. The ATLAS and CMS high-level trigger systems operate under event-processing latency budgets of tens of milliseconds per event~\cite{atlas_hllhc_comp,cms_hllhc_comp}, a constraint that exact quadratic neighbourhood search cannot satisfy at $N \sim 10^5$ given current GPU throughput. An additional consequence is that the intermediate $N \times N$ distance matrix required by exact search exceeds single-GPU memory capacity at full HL-LHC scale, so the computation and memory bottlenecks are inseparable.

Replacing exact $k$-NN search with approximate nearest-neighbour methods such as FAISS~\cite{8733051} can reduce per-layer graph construction from $\mathcal{O}(N^2)$ toward $\mathcal{O}(N \log N)$ in both time and peak memory, while potentially preserving high neighbour quality at tracking-relevant distance scales. Since the remaining stages already scale as $\mathcal{O}(Nk)$, this substitution could reduce the overall per-layer complexity to $\mathcal{O}(N \log N)$,
providing a practical route toward HL-LHC-scale deployment.

\subsection{Limitations and open questions}

The present study is limited to simulated benchmarks. Real detector operation introduces additional effects, including detector inefficiencies, material interactions, noise, and realistic pile-up distributions that are not fully represented in the datasets considered here. Validation under such conditions, together with evaluation at substantially larger event sizes, is necessary before drawing conclusions about practical deployment.

\section{Conclusion}

We introduced an Evolving-Geometry Transformer for charged-particle track reconstruction, combining layer-wise neighbourhood refinement with geometry-aware attention. The architecture allows the interaction topology to adapt as hit representations evolve, while relative hit geometry directly modulates pairwise feature interactions.

Across five benchmarks ranging from simple linear tracks to high-density TrackML-scale events, EGT achieves best-in-class FitAccuracy on three of five settings and competitive performance on the remainder. The most significant gains appear in geometrically complex and high-density scenarios, which are representative of conditions expected in HL-LHC operation, indicating that the evolving neighbourhood mechanism and geometry-aware attention are particularly beneficial in challenging relational settings. This trend is consistent with the role of evolving topology when static geometric proximity becomes less effective for distinguishing track membership.

The ablation results provide the clearest evidence for the role of topology refinement. Replacing the evolving graph with a static topology reduces FitAccuracy from $0.9388$ to $0.7864$ and the Perfects rate from $0.81$ to $0.29$, whereas removing geometry-conditioned bias reduces FitAccuracy to $0.9029$ and Perfects to $0.73$. This pronounced asymmetry indicates that adaptive neighbourhood refinement is the dominant contribution to complete-track recovery in the evaluated setting, with relative geometry providing a complementary improvement.

Taken together, these results support a view of dense particle tracking as an iterative relational inference problem: as trajectory-level representations become more informative, the interaction structure between hits should be refined accordingly. Extending this principle to full HL-LHC event sizes will require more scalable neighbourhood construction and validation under realistic detector and pile-up conditions.

\appendix

\section{Implementation details}\label{app:impl}

We adopt the same loss function as EncReg~\cite{caron2025trackformers} and train all models using the Adam optimizer. The batch size is 64 for all datasets except TrackML 200--500, where it is reduced to 8 due to memory constraints. Learning rates are in the range $[8\times10^{-4},\, 1.1\times10^{-3}]$, with cosine annealing applied to the two REDVID datasets and a fixed rate for the remainder. 

Table~\ref{tab:training_details} summarises the dataset-specific training configurations used in all experiments.

\begin{table}[h]
\centering
\small

\setlength{\tabcolsep}{4pt}
\begin{tabular}{lccc}
\toprule
Dataset & Batch & LR & Epochs \\
\midrule
REDVID -- 10--50 linear   & 64 & $8{\times}10^{-4}$   & 180 \\
REDVID -- 10--50 helical  & 64 & $1.1{\times}10^{-3}$ & 120 \\
REDVID -- 50--100 helical & 64 & $1.0{\times}10^{-3}$ & 100 \\
TrackML -- 10--50         & 64 & $1.0{\times}10^{-3}$ & 100 \\
TrackML -- 200--500       & 8  & $1.0{\times}10^{-3}$ & 100 \\
\bottomrule
\end{tabular}

\smallskip
\caption{Dataset-specific training configurations.}
\label{tab:training_details}

\noindent Cosine annealing is applied for REDVID 10--50 linear and REDVID 10--50 helical
($T_{\max}$ equal to the number of training epochs); all other datasets
use a fixed learning rate.
\end{table}

We use dataset-specific architectural configurations to balance model capacity with event complexity. For the first four datasets, the model consists of 6 geometry-aware graph transformer blocks with embedding size 32 and hidden dimension 128. For the fifth dataset (TrackML 200--500), we use 7 blocks, embedding size 128, and hidden dimension 256.

The neighbourhood size for evolving $k$NN graph construction is set to $k=64$ for all datasets except REDVID 50--100 helical, where $k=128$.

All models use 4 attention heads with a per-head dimension of 8, followed by an output projection back to the full hidden dimension. The relation encoder $\psi$ is implemented as a two-layer MLP with a single ReLU activation between the linear layers; no dropout is applied within this module.

For HDBSCAN post-processing, we independently tune the minimum cluster size over $\{3, 4, 5, 6\}$ and the minimum number of samples over $\{1, 2, 3, 4, 5\}$ for each dataset on the validation set.

All experiments are conducted on a single NVIDIA A100 80\,GB GPU.

\subsection{Inference latency measurement protocol}

Instance-level inference latency is measured with each test event processed independently (batch size~$=1$); the first event is discarded as warmup and means are taken over the held-out test split (15\%, seed~37). GPU time is measured with \texttt{cuda.Event} and covers only the neural-network forward pass; CPU time uses \texttt{process\_time\_ns} and includes post-processing on the host, comprising hit unpadding and HDBSCAN track clustering. EncReg latencies are taken from TrackFormers~\cite{caron2025trackformers} (NVIDIA A100 40\,GB, 18 CPU cores); all EGT numbers are measured in this work on an NVIDIA A100 80\,GB PCIe GPU.

For a fair latency comparison, an inference-only implementation is deployed that preserves the trained weights but replaces the dense $(N{\times}N)$ attention mask with sparse $k$-NN aggregation: attention is restricted to $k$ neighbours per query with keys and values gathered by index rather than materialising a full attention matrix; only non-padded hits act as attention queries, avoiding redundant computation on approximately 70\% of the padded tensor in the TrackML 200--500 setting (${\sim}1\,695$ valid hits out of 4\,717 padded slots per event); and when padding dominates ($N_{\mathrm{valid}}/N_{\mathrm{pad}} < 0.85$), $k$-NN graphs are built on valid hits only with indices remapped using GPU \texttt{cdist} rather than a full padded distance matrix. Reconstruction performance is unchanged relative to the training-time forward pass, with mean event-level FitAccuracy differing by less than $10^{-4}$ across all five datasets.

\subsection{GPU profiling at TrackML 200--500}

Profiling is performed with CUDA events at the highest complexity setting (TrackML 200--500, $k{=}64$, seven encoder layers, ${\sim}1\,695$ valid hits per event). The edge MLP, evaluated once per query--neighbour pair, accounts for approximately 64\% of total GPU forward time; the remainder is distributed over sparse attention, per-layer $k$-NN reconstruction, and feed-forward blocks. GPU inference is therefore dominated by the per-edge bias network rather than by global attention or graph construction, identifying the edge MLP as the primary target for future latency reduction.

\section{HDBSCAN hyperparameters}\label{app:hdbscan}

\begin{figure}[t]
    \centering
    \includegraphics[width=0.49\columnwidth]{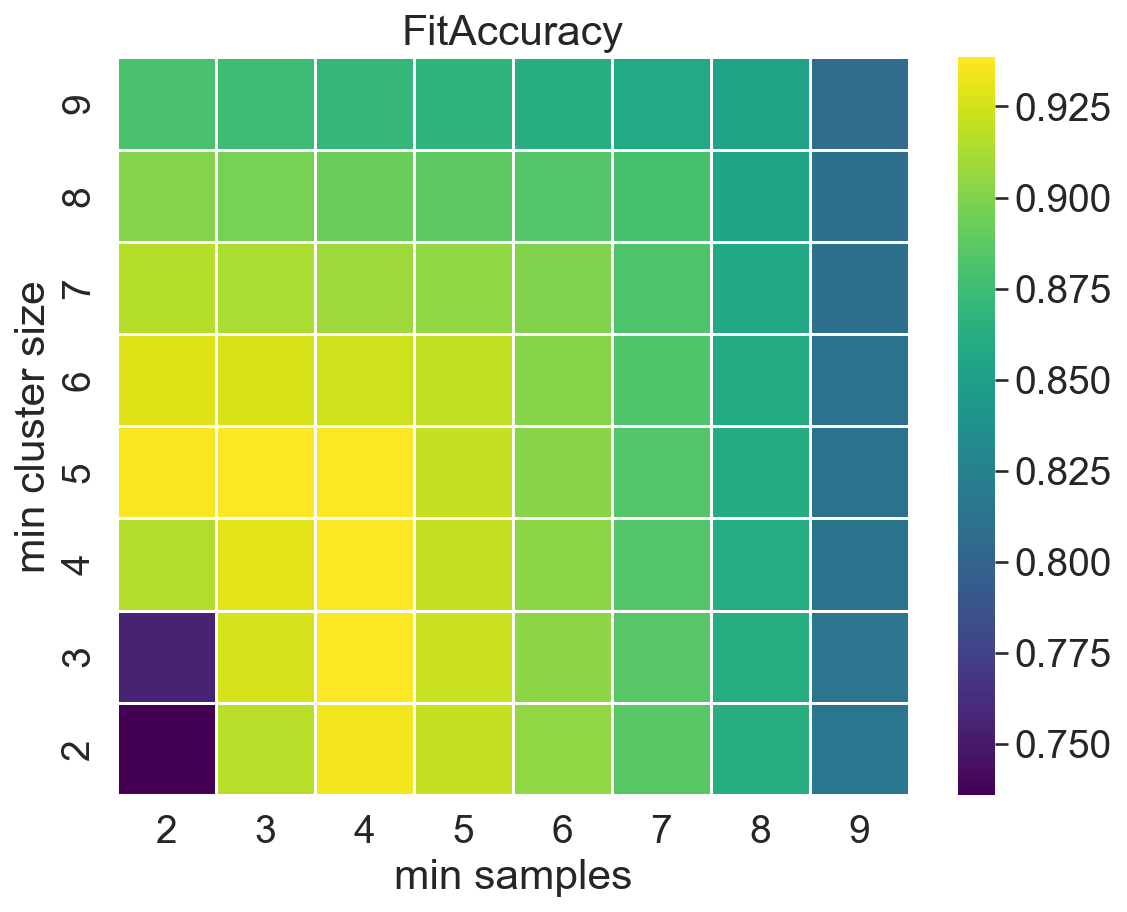}
    \hfill
    \includegraphics[width=0.49\columnwidth]{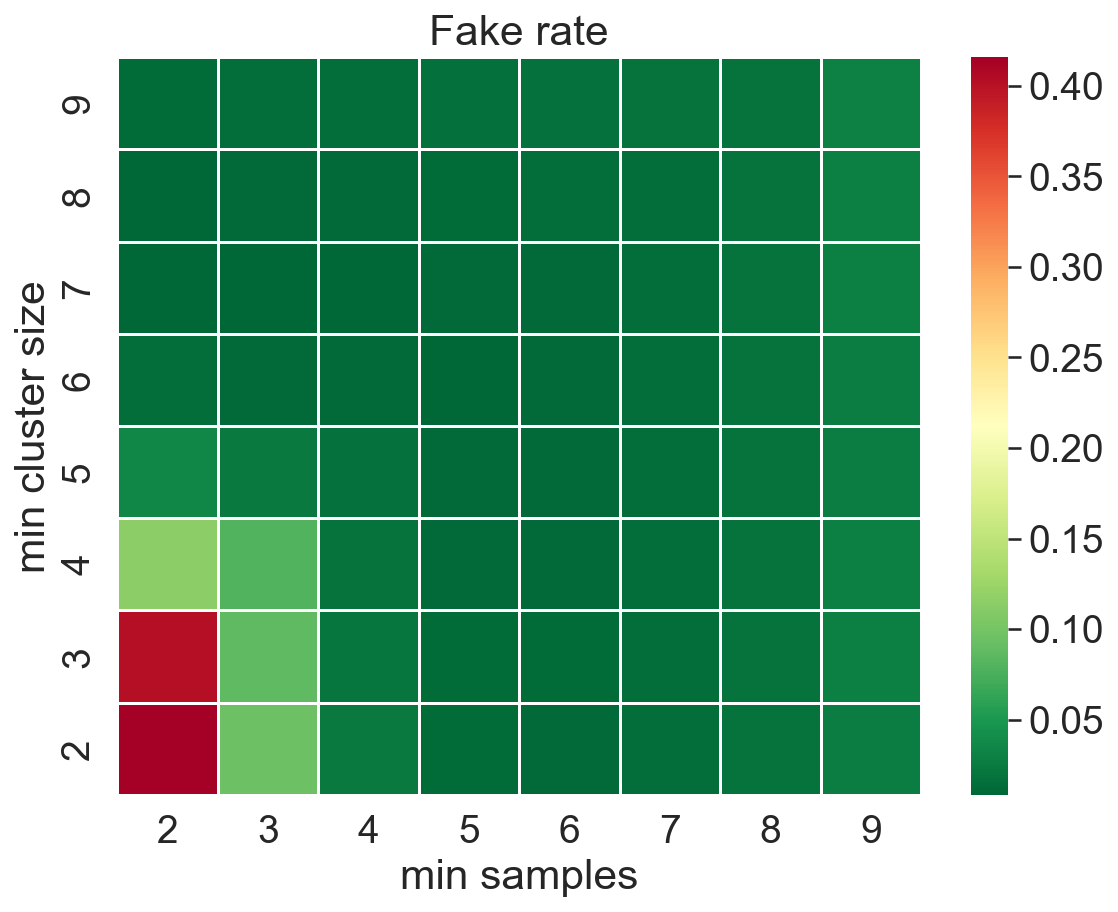}
    \caption{HDBSCAN post-processing parameter sensitivity on the TrackML 10--50 validation set.
    The left panel shows FitAccuracy, while the right panel shows the fake track rate.
    \texttt{min\_cluster\_size}$\,\in\{2,3\}$ and \texttt{min\_samples}$\,=2$ exhibit fake rates above $40\%$, illustrating the rapid degradation that occurs for overly permissive clustering settings. The selected configuration (\texttt{min\_cluster\_size}$\,=5$, \texttt{min\_samples}$\,=3$) lies at the intersection of the high-accuracy plateau and the low-fake-rate region.}
    \label{fig:hdbscan_sweep}
\end{figure}

Following per-hit predictions of $(\theta,\sin\phi,\cos\phi,q)$, HDBSCAN clusters hits in this four-dimensional parameter space into track candidates. The two hyperparameters control the minimum cluster size and the local density required for forming track candidates, and their effects can therefore be related to track multiplicity and cluster stability. For the sensitivity study in figure~\ref{fig:hdbscan_sweep}, the \texttt{min\_cluster\_size} range is extended to $2$--$9$ on the TrackML 10--50 validation set, beyond the range used for hyperparameter selection.

\texttt{min\_cluster\_size} sets the minimum number of predicted hits required to form a track candidate. Small values allow short accidental groups in the predicted parameter space to be accepted as clusters, whereas larger values suppress such groups at the cost of potentially rejecting trajectories represented by fewer hits. The validation sweep therefore determines an operating point that balances FitAccuracy and fake-track rate. The sharp increase in fake rate at small \texttt{min\_cluster\_size} values indicates that these settings admit a large number of accidental clusters in the predicted parameter space.

\texttt{min\_samples} controls the local density required for a prediction to be treated as a core point by HDBSCAN. Small values make cluster formation more permissive and can increase accidental merging, whereas large values can reject sparse but otherwise valid clusters. The selected configuration, \texttt{min\_cluster\_size}$=5$ and \texttt{min\_samples}$=3$, lies in a region with simultaneously high FitAccuracy and low fake-track rate in figure~\ref{fig:hdbscan_sweep}.

\acknowledgments

 The authors were partially supported by the National Natural Science Foundation of China (Grants No. 62576221, 62002333), the Guangdong Provincial Natural Science Foundation (No. 2025A1515010288), and the Shenzhen International Quantum Academy (Grant No. SIQA2024KFKT01).
 
 During the preparation of this work, the authors used OpenAI's ChatGPT and Codex to assist with English-language editing, implementation planning, and code generation. The research objectives and final methodological decisions were determined by the authors. AI-generated implementation suggestions and code were reviewed by the authors and verified for consistency with the intended methodology. The authors take full responsibility for the scientific content, implementation, results, and conclusions of this work.

\paragraph*{Data Availability Statement.}

No new datasets were generated in this study. The datasets analysed in this work are publicly available through the TrackFormers dataset repository on Zenodo at \url{https://doi.org/10.5281/zenodo.14386134}.

\paragraph*{Code Availability Statement.}

The implementation builds in part on the publicly available TrackFormers codebase. The EGT-specific code developed in this study is available from the corresponding author upon reasonable request.

\bibliographystyle{JHEP}
\bibliography{biblio}

\end{document}